# Chiral effect in plane isotropic micropolar elasticity and its application to chiral lattices

X.N. Liu[1,2], G.L. Huang[2*] and G.K. Hu[1*]

[1]*Key Laboratory of Dynamics and Control of Flight Vehicle, Ministry of Education, School of Aerospace Engineering, Beijing Institute of Technology, Beijing 100081, China*

[2]*Department of Systems Engineering, University of Arkansas at Little Rock, Little Rock, Arkansas 72204*

**Abstract**

In continuum mechanics, the non-centrosymmetric micropolar theory is usually used to capture the chirality inherent in materials. However when reduced to a two dimensional (2D) isotropic problem, the resulting model becomes non-chiral. Therefore, influence of the chiral effect cannot be properly characterized by existing theories for 2D chiral solids. To circumvent this difficulty, based on reinterpretation of isotropic tensors in a 2D case, we propose a continuum theory to model the chiral effect for 2D isotropic chiral solids. A single material parameter related to chirality is introduced to characterize the coupling between the bulk deformation and the internal rotation which is a fundamental feature of 2D chiral solids. Coherently, the proposed continuum theory is also derived for a triangular chiral lattice from a homogenization procedure, from which the effective material constants of the lattice are analytically determined. The unique behavior in the chiral lattice is demonstrated through the analyses of a static tension problem and a plane wave propagation problem. The results, which cannot be predicted by the non-chiral model, are validated by the exact solution of the discrete model.

---

[*] Corresponding author: glhuang@ualr.edu; hugeng@bit.edu.cn;

# 1. Introduction

An object is said to be chiral, or with handedness, if it cannot be superposed to its mirror image (Lord Kelvin, 1904). Chirality is encountered in many branches of science, including physics, biology, chemistry and optics. A chiral material should be described by an adequate constitutive equation with handedness in order to characterize the distinct feature of such material. In continuum mechanics, chirality is considered in the context of generalized elasticity, e.g. micropolar (Cosserat) theory (Cosserat E. and Cosserat F., 1909; Eringen, 1966). A general isotropic chiral (also known as non-centrosymmetric, acentric or hemitropic) micropolar theory introduces three additional material constants compared to the non-chiral theory, the additional material parameters change their signs according to the handedness of the microstructure to represent the effect of chirality (Nowacki, 1986; Lakes and Benedict, 1982; Lakes, 2001; Natroshvili and Stratis, 2006a,b; Joumaa and Ostoja-Starzewski, 2011). This theory provides an efficient tool for modeling the chiral effect presented in materials and structures, e.g., loading transfer in carbon nanotubes and chiral rods (Chandraseker and Mukherjee, 2006; Chandraseker et al, 2009; Ieşan, 2010), mechanics of bone (Lakes et al, 1983), chirality transfer in nanomaterials (Wang et al, 2011) and wave propagation in chiral solids (Lakhtakia et al, 1988; Ro, 1999; Khurana and Tomar, 2009). However, when this theory is applied to a planar isotropic case, e.g. a triangular chiral lattice, the variables describing the chiral effect disappear and the resulting theory becomes basically non-chiral (Spadoni and Ruzzene, 2012). Therefore the basic characteristic of a planar chiral solid cannot be properly modeled by the existing theory.

On the other hand, lattice structures can be homogenized as micropolar continuum media (Bazant and Christensen, 1972; Chen et al, 1998; Kumar and McDowell, 2004), the homogenized material constants are derived directly from their microstructures. This provides a useful tool to explain the observed size effect in lattice structures. Chiral lattice structure was also proposed by Prall and Lakes (1996) to achieve a material with negative Poisson's ratio (Lakes, 1987). Among the candidates of these so-called auxetic materials (Yang et al, 2004), the triangular chiral lattice is the mostly investigated one since it is isotropic and the geometric pattern can be controlled by a single continuously varying topological parameter. Its unique mechanical behavior was examined by many researchers for both static (Alderson et al, 2010; Dirrenberger et al, 2011; Spadoni and Ruzzene, 2012) and dynamic (Spadoni et al, 2009) loading cases with a number of targeted applications. The chiral material was also used in designing elastic metamaterials with the negative effective bulk modulus (Liu et al, 2011). Recently Spadoni and Ruzzene (2012) proposed a self-consistent homogenization scheme for a 2D chiral lattice in the framework of the micropolar theory in order to clarify the indeterminacy of the effective shear modulus. However, since the non-centrosymmetric isotropic micropolar model becomes non-chiral when applied to a planar problem (Spadoni and Ruzzene, 2012), the developed homogenization method in this framework cannot characterize the chiral effect inherent in the material. Therefore we are encountering a challenge problem: for planar isotropic chiral materials, e.g. triangular chiral lattices, we do not have a solid theory either in continuum formulation or in homogenization method to characterize their chiral effects.

The objective of the paper is to propose a continuum model to capture the chiral effect for planar isotropic chiral solids, and the corresponding effective material constants will be derived for a planar triangular chiral lattice. The theory will also be illustrated through some examples to demonstrate its necessity and consistency in characterizing the chiral effect. The manuscript is organized as follows: in Section 2, a new constitutive relation for a 2D isotropic chiral solid is proposed based on a continuum formulation. In Section 3, a triangular chiral lattice structure is homogenized in the framework of the proposed theory and the effective material constants are derived. In Section 4, a tension and plane wave propagation problems are examined for a planar chiral lattice by the proposed theory. In Section 5 the main result of this work is concluded.

## 2. Planar isotropic micropolar model with chirality

Characterization of material chirality is closely related to the concept of pseudo (or axial) tensors, they alternate the sign with a mirror reflecting transformation or the handedness change of the underlying coordinate system, and ordinary (or polar) tensors are not affected by such actions (Borisenko and Tarapov, 1979). Both types of tensors coexist in various elastic formulations, but strain energy density must be independent of handedness.

Classical elasticity theory excludes chirality (Lakes and Benedict, 1982), since in the energy density

$$w = \frac{1}{2}\varepsilon_{ij}C_{ijkl}\varepsilon_{kl}, \tag{1}$$

the strain $\varepsilon$ is an ordinary tensor. To maintain $w$ as an ordinary scalar, the elastic tensor $C$ cannot be pseudo. So to include chirality, micropolar theory (Eringen, 1999) is considered in this paper. In the micropolar theory, rotational degree of freedom (DOF) $\phi_i$ is introduced in addition to the displacement $u_i$ on a material point. The strain and curvature play as deformation measures

$$\varepsilon_{kl} = u_{l,k} + e_{lkm}\phi_m, \tag{2a}$$

$$\kappa_{kl} = \phi_{k,l}, \tag{2b}$$

and the balance of stress $\sigma_{ji}$ and couple stress $m_{ji}$ are governed by

$$\sigma_{ji,j} = \rho\partial^2 u_i/\partial t^2, \tag{3a}$$

$$m_{ji,j} + e_{ikl}\sigma_{kl} = j\partial^2 \phi_i/\partial t^2, \tag{3b}$$

where $e_{ijk}$ is the Levi-Civita tensor, $\rho$ and $j$ are the density and micro-inertia, respectively. In this paper, the Einstein's summation convention is used and the comma in subscript denotes partial differentiation with respect to spatial coordinates. The strain energy density for a general linear elastic micropolar media is expressed as a quadratic form in terms of asymmetric strain and curvature tensors

$$w = \frac{1}{2}\varepsilon_{ij}C_{ijkl}\varepsilon_{kl} + \frac{1}{2}\kappa_{ij}D_{ijkl}\kappa_{kl} + \varepsilon_{ij}B_{ijkl}\kappa_{kl}, \tag{4}$$

where $C$, $D$ and $B$ are elastic tensors of rank four. Then the constitutive relation can be derived as

$$\sigma_{ij} = C_{ijkl}\varepsilon_{kl} + B_{ijkl}\phi_{k,l}, \tag{5a}$$

$$m_{ij} = B_{ijkl}\varepsilon_{kl} + D_{ijkl}\phi_{k,l}. \tag{5b}$$

It should be noted that the microrotation vector $\phi_i$ and curvature tensor $\kappa_{ij}$ are pseudo, thus in Eq.(4), $B_{ijkl}$ must be a pseudo tensor and thus represent the chirality. A micropolar solid with $B_{ijkl} \neq 0$ is usually referred as non-centrosymmetric. Let us consider the isotropic case and focus only the chiral part of Eq.(5). A general isotropic tensor of rank four takes the form as

$$B_{ijkl} = B_1 \delta_{ij} \delta_{kl} + B_2 \delta_{ik} \delta_{jl} + B_3 \delta_{il} \delta_{jk}, \tag{6}$$

where $\delta_{jk}$ is Kronecker delta. The chiral part of Eq.(5) then reads

$$\sigma_{ij} = B_1 \delta_{ij} \phi_{k,k} + B_2 \phi_{i,j} + B_3 \phi_{j,i}, \tag{7a}$$

$$m_{ij} = B_1 \delta_{ij} \varepsilon_{kk} + B_2 \varepsilon_{ij} + B_3 \varepsilon_{ji}. \tag{7b}$$

A planar micropolar problem in $x_1 - x_2$ plane is defined by $u_3 = \phi_1 = \phi_2 = \partial/\partial x_3 = 0$, while the non zero quantities are $u_\alpha$, $\phi_3$, $\phi_{3,\alpha}$, $\varepsilon_{\alpha\beta}$, $\sigma_{\alpha\beta}$ and $m_{\alpha 3}$, respectively, with Greek subscripts ranging from 1 to 2. Specialized to the 2D case, it is easy to verify that Eq.(7) is trivial, as a result, chirality represented by the isotropic $B_{ijkl}$ disappears. However, for 2D isotropic chiral solids, e.g., planar triangular chiral lattices, chirality is a basic feature of such structures and should be characterized by a correct constitutive modeling. Therefore there should be something missing in Eqs. (5) and (6) when reduced to the 2D case.

To circumvent this problem, we first discuss some basic properties of isotropic tensors. The basic forms of isotropic tensors of rank 0, 2 and 3 are just the scalar, Kronecker delta $\delta_{jk}$ and Levi-Civita tensor $e_{ijk}$, respectively. A vector which is rank one cannot be isotropic, and $e_{ijk}$ is a pseudo tensor. Any isotropic tensor with rank greater than three can be constructed by scalar, $\delta_{jk}$ and $e_{ijk}$, just as Eq.(6) for an isotropic tensor of rank four. However, in the 2D case, the Levi-Civita tensor $e_{ijk}$ is restricted to the form $e_{ijk} \equiv e_{3\alpha\beta}$, which is in-plane isotropic and equivalent to an isotropic tensor of rank two. For the same reason, there is no in-plane third order isotropic tensor in the current 2D situation. The isotropic $B$ tensor also vanishes, since the energy density of the 2D case can be rewritten as

$$w = \frac{1}{2}\varepsilon_{\alpha\beta}C_{\alpha\beta\gamma\rho}\varepsilon_{\gamma\rho} + \frac{1}{2}\phi_{3,\alpha}D_{\alpha\beta}\phi_{3,\beta} + \varepsilon_{\alpha\beta}B_{\alpha\beta\gamma}\phi_{3,\gamma},\tag{8}$$

where $B$ reduces to the tensor of rank three and cannot be isotropic except zero. However, with $e_{3\alpha\beta}$ and $\delta_{\alpha\beta}$, we thus have more choices to construct a 2D isotropic tensor of rank four. In particular, we have

$$\bar{C}_{\alpha\beta\gamma\rho} = \bar{C}_1\delta_{\alpha\beta}\delta_{\gamma\rho} + \bar{C}_2\delta_{\alpha\gamma}\delta_{\beta\rho} + \bar{C}_3\delta_{\alpha\rho}\delta_{\beta\gamma},\tag{9a}$$

$$\tilde{C}_{\alpha\beta\gamma\rho} = \tilde{C}_1\delta_{\alpha\beta}e_{3\gamma\rho} + \tilde{C}_2\delta_{\gamma\rho}e_{3\alpha\beta} + \tilde{C}_3\delta_{\alpha\gamma}e_{3\beta\rho} + \tilde{C}_4\delta_{\beta\rho}e_{3\alpha\gamma} + \tilde{C}_5\delta_{\alpha\rho}e_{3\beta\gamma} + \tilde{C}_6\delta_{\beta\gamma}e_{3\alpha\rho},\tag{9b}$$

$$\hat{C}_{\alpha\beta\gamma\rho} = \hat{C}_1 e_{3\alpha\beta}e_{3\gamma\rho} + \hat{C}_2 e_{3\alpha\gamma}e_{3\beta\rho} + \hat{C}_3 e_{3\alpha\rho}e_{3\beta\gamma}.\tag{9c}$$

Eq.(9a) is just the 2D version of Eq.(6). By utilizing the identity $e_{3\alpha\beta}e_{3\gamma\rho} = \delta_{\alpha\gamma}\delta_{\beta\rho} - \delta_{\alpha\rho}\delta_{\beta\gamma}$, Eq.(9c) is found to take the same form as Eq.(9a), this is not surprising since $\hat{C}_{\alpha\beta\gamma\rho}$ is the product of two pseudo tensors, hence it is an ordinary tensor. In summary, we conclude that for a 2D micropoar problem the generic form of an isotropic tensor of rank four can be given by

$$C_{\alpha\beta\gamma\rho} = \bar{C}_{\alpha\beta\gamma\rho} + \tilde{C}_{\alpha\beta\gamma\rho}.\tag{10}$$

Reexamining the energy density in Eq.(8) with the help of Eq.(10) and $D_{\alpha\beta} = D_1\delta_{\alpha\beta}$, we have

$$w = \frac{1}{2}\varepsilon_{\alpha\beta}\bar{C}_{\alpha\beta\gamma\rho}\varepsilon_{\gamma\rho} + (\tilde{C}_1 + \tilde{C}_2)\varepsilon_{\alpha\alpha}e_{3\gamma\rho}\varepsilon_{\gamma\rho} + \frac{1}{2}D_1\phi_{3,\alpha}\delta_{\alpha\beta}\phi_{3,\beta}.\tag{11}$$

Introducing the Lame's constant $\lambda$, $\mu$, antisymmetric shear modulus $\kappa$, higher order modulus $\gamma$, and a single chiral parameter $2A \equiv \tilde{C}_1 + \tilde{C}_2$, the in-plane isotropic micropolar elastic tensors with chirality are thus written as

$$\bar{C}_{\alpha\beta\gamma\rho} = \lambda\delta_{\alpha\beta}\delta_{\gamma\rho} + (\mu+\kappa)\delta_{\alpha\gamma}\delta_{\beta\rho} + (\mu-\kappa)\delta_{\alpha\rho}\delta_{\beta\gamma},\tag{12a}$$

$$\tilde{C}_{\alpha\beta\gamma\rho} = A(\delta_{\alpha\beta}e_{3\gamma\rho} + \delta_{\gamma\rho}e_{3\alpha\beta}),\tag{12b}$$

$$\bar{D}_{\alpha\beta} = \gamma\delta_{\alpha\beta}.\tag{12c}$$

Note that Eq.(12b) has a symmetric form conforming with the requirement of major symmetry for the $C$ tensor. The constitutive equation then becomes

$$\sigma_{\alpha\beta} = \lambda\delta_{\alpha\beta}\varepsilon_{\rho\rho} + (\mu+\kappa)\varepsilon_{\alpha\beta} + (\mu-\kappa)\varepsilon_{\beta\alpha} + A\delta_{\alpha\beta}e_{3\gamma\rho}\varepsilon_{\gamma\rho} + Ae_{3\alpha\beta}\varepsilon_{\rho\rho},\tag{13}$$

$$m_{\alpha 3} = \gamma \phi_{3,\alpha}.$$

It is interesting to note that the pseudo tensor $\tilde{C}_{\alpha\beta\gamma}$ representing the chirality relates to both the normal stress and normal strain, this is different from the $B_{ijkl}$ tensor for a 3D case. The physical meaning of $\tilde{C}_{\alpha\beta\gamma}$ is however very clear. Consider the relevant part in Eq.(11)

$$A(\delta_{\alpha\beta}e_{3\gamma\rho} + \delta_{\gamma\rho}e_{3\alpha\beta})\varepsilon_{\alpha\beta}\varepsilon_{\gamma\rho} = 2A\varepsilon_{\rho\rho}(e_{3\alpha\beta}\varepsilon_{\alpha\beta}), \tag{14}$$

the spherical strain $\varepsilon_{\rho\rho}$ represents the bulk deformation at a material point, it is obviously independent of the handedness of the frame. On the other hand, $e_{3\alpha\beta}\varepsilon_{\alpha\beta} = -2(\phi_3 - \psi_3)$ is the *pure* rotation of a point, with $\psi_3 = e_{3\alpha\beta}u_{\beta,\alpha}/2$ denoting the macro rigid rotation, therefore, it is an axial quantity depending on the handedness. This chiral term in the energy density clearly demonstrates that a pure rotation can produce shrink or dilatation of the material, and vice versa. This mechanism derived from a continuum formulation explains correctly the behavior of a real 2D chiral structure( e.g. the triangular chiral lattice schematically depicted in Fig. 1), which will be discussed in detail in the following section. This is also the unique mechanism of the chiral lattice to produce the negative Poisson's ratio.

The constitutive law of Eq. (13) can be rearranged in a matrix form as

$$\begin{Bmatrix} \sigma_{11} \\ \sigma_{22} \\ \sigma_{12} \\ \sigma_{21} \\ m_{13} \\ m_{23} \end{Bmatrix} = \begin{bmatrix} 2\mu+\lambda & \lambda & -A & A & 0 & 0 \\ \lambda & 2\mu+\lambda & -A & A & 0 & 0 \\ -A & -A & \mu+\kappa & \mu-\kappa & 0 & 0 \\ A & A & \mu-\kappa & \mu+\kappa & 0 & 0 \\ 0 & 0 & 0 & 0 & \gamma & 0 \\ 0 & 0 & 0 & 0 & 0 & \gamma \end{bmatrix} \begin{Bmatrix} u_{1,1} \\ u_{2,2} \\ u_{2,1}-\phi \\ u_{1,2}+\phi \\ \phi_{,1} \\ \phi_{,2} \end{Bmatrix}. \tag{15}$$

It has four classical micropolar elastic constants and a new parameter $A$ characterizing the chiral effect. When the handedness of the material pattern is flipped over, the chiral constant $A$ should reverse its sign to maintain the invariance of the strain energy density, and the other constants remain unchanged:

$$\begin{Bmatrix} \sigma_{11} \\ \sigma_{22} \\ \sigma_{12} \\ \sigma_{21} \\ m_{13} \\ m_{23} \end{Bmatrix} = \begin{bmatrix} 2\mu+\lambda & \lambda & A & -A & 0 & 0 \\ \lambda & 2\mu+\lambda & A & -A & 0 & 0 \\ A & A & \mu+\kappa & \mu-\kappa & 0 & 0 \\ -A & -A & \mu-\kappa & \mu+\kappa & 0 & 0 \\ 0 & 0 & 0 & 0 & \gamma & 0 \\ 0 & 0 & 0 & 0 & 0 & \gamma \end{bmatrix} \begin{Bmatrix} u_{1,1} \\ u_{2,2} \\ u_{2,1}-\phi \\ u_{1,2}+\phi \\ \phi_{,1} \\ \phi_{,2} \end{Bmatrix} \quad (16)$$

Finally, with the proposed constitutive equation, the governing equations expressed in the displacements $u$, $v$ and the microrotation $\phi \equiv \phi_3$ read

$$\rho \frac{\partial^2 u}{\partial t^2} = (\lambda+2\mu)u_{,xx} + (\mu+\kappa)u_{,yy} + (\lambda+\mu-\kappa)v_{,xy} + 2\kappa\phi_{,y} - A(v_{,xx} - v_{,yy} - 2u_{,xy} - 2\phi_{,x}), \quad (17a)$$

$$\rho \frac{\partial^2 v}{\partial t^2} = (\mu+\kappa)v_{,xx} + (\lambda+2\mu)v_{,yy} + (\lambda+\mu-\kappa)u_{,xy} - 2\kappa\phi_{,x} - A(u_{,xx} - u_{,yy} + 2v_{,xy} - 2\phi_{,y}), \quad (17b)$$

$$j\frac{\partial^2 \phi}{\partial t^2} = \gamma(\phi_{,xx} + \phi_{,yy}) - 4\kappa\phi + 2\kappa(v_{,x} - u_{,y}) - 2A(u_{,x} + v_{,y}). \quad (17c)$$

The form of Eqs. (15-17) looks like those of an anisotropic medium, however, they are fundamentally different and cannot be covered by even the most general anisotropy without chirality, since the parameter $A$ and its sign form a unique pattern in the constitutive matrix. They are indeed in-plane isotropic guaranteed by Eq. (12). It should be mentioned that, for a 2D isotropic micropolar solid, Eq.(12) is the only possible form to include chiralty.

The constraint conditions on the five material constants can be obtained by imposing positive definiteness of the strain energy density $w$. It is convenient to decompose the strain tensor into hydrostatic, deviatoric symmetric and antisymmetric parts as (Liu and Hu, 2005)

$$\varepsilon_{\alpha\beta} = \delta_{\alpha\beta}\bar{\varepsilon} + \varepsilon^d_{(\alpha\beta)} + \varepsilon_{<\alpha\beta>}, \quad (18)$$

where

$$\bar{\varepsilon} = \frac{1}{2}\varepsilon_{\alpha\alpha}, \quad (19a)$$

$$\varepsilon^d_{(\alpha\beta)} = \frac{1}{2}(\varepsilon_{\alpha\beta} + \varepsilon_{\beta\alpha}) - \delta_{\alpha\beta}\bar{\varepsilon}, \quad (19b)$$

$$\varepsilon_{<\alpha\beta>} = \frac{1}{2}(\varepsilon_{\alpha\beta} - \varepsilon_{\beta\alpha}) = e_{\beta\alpha 3}(\phi-\psi). \quad (19c)$$

Substituting Eqs.(18) and (12) into (11) yields

$$w = 2\bar{\kappa}\,\bar{\varepsilon}^2 + 2\mu\varepsilon^d_{(\alpha\beta)}\varepsilon^d_{(\alpha\beta)} + 2\kappa(\phi-\psi)^2 + 4A\bar{\varepsilon}(\phi-\psi) + \frac{1}{2}\gamma\phi_{,\alpha}\phi_{,\alpha}, \tag{20}$$

where $\bar{\kappa} = \lambda + \mu$ is defined as the 2D bulk (area) modulus. This equation yields the necessary and sufficient conditions for the positive definiteness of $w$. Besides the four conditions for a classical micropolar medium given in the literature

$$\bar{\kappa} = \lambda + \mu > 0, \quad \mu > 0, \quad \kappa > 0, \quad \gamma > 0, \tag{21}$$

an additional condition

$$A^2 < (\lambda + \mu)\kappa, \tag{22}$$

is imposed on the chiral constant $A$. It can be either positive or negative, but its absolute value is bounded.

## 3. Homogenization for a 2D triangular chiral lattice

In this section, we will examine a 2D triangular chiral lattice, and analytically derive the five material constants proposed in the section 2 by a homogenization method.

*3.1 Description of Geometry*

The geometry of the chiral lattice is shown in Fig. 2. The structural layout consists of circles of radius $r$ linked by straight ligaments of equal length $L$. The ligaments are required to be tangential to the circles and the angle between adjacent ligaments is $\pi/3$. The circles are arranged as a triangular lattice with the lattice constant, i.e. the distance between circle centers, $a$. The angle between the line connecting the circle centers and the ligament is defined as $\beta$. The geometry parameters are connected by the following relations:

$$\cos\beta = \frac{L}{a}, \qquad \sin\beta = \frac{2r}{a}. \tag{23}$$

A quadrilateral unit cell is used for the analytical development, as bounded by the dashed parallelogram in Fig. 2. The tessellation of the unit cell along the lattice vectors

$$\begin{aligned}\mathbf{a}_1 &= a\mathbf{e}_x \\ \mathbf{a}_2 &= -\frac{1}{2}a\mathbf{e}_x + \frac{\sqrt{3}}{2}a\mathbf{e}_y\end{aligned} \tag{24}$$

generates the whole lattice. The wall thickness of the circles and the ligaments are denoted as $t_c$ and $t_b$, respectively.

Among the parameters, $\beta$ is denoted as a topology parameter (Prall and Lakes, 1996; Spadoni et al., 2009). When $\beta \to 0$, the circles shrink to dots and a traditional triangular lattice is obtained. When $\beta \to \pi/2$ the ligaments vanish and a lattice of packed circles is recovered. It should be noted that the chiral effect disappears in these two extreme cases. The parameter $\beta$ affects not only the topology but also the mechanical behavior of the lattice. As pointed by Spadoni and Ruzzene (2012), the variation of $\beta$ monitors the transition from bending dominated behavior to axially dominated behavior.

Finally, as shown in Fig. 3, we adopt the sign convention of $\beta$ according to the relative orientation of the ligament and the link of circle centers. When $\beta$ reverses its sign, the lattice is flipped over and a complementary pattern with a reversed handedness is implied. This operation cannot be achieved by an in-plane rotation due to its chiral nature.

*3.2 Determination of constitutive constants*

To formulate the problem analytically, the circles of the chiral lattice are assumed to be rigid. Further, the ligament is assumed to be massless, and the mass and rotation inertia of the rigid circle are denoted as $m$ and $J$, respectively. However, a numerical procedure (Spadoni and Ruzzene, 2012) can be utilized to deal with the case of deformable circles and distributed mass density.

Let $u_i$, $v_i$ and $\phi_i$ denote the displacement and rotation DOFs at the center of the rigid circle $i$, expressing in a vector form,

$$\mathbf{u}_i = \{u_i \quad v_i \quad \phi_i\}^T. \tag{25}$$

Motion of the center of the rigid circle is enforced to the beam end by the motion $\tilde{\mathbf{u}}_i$ on its

circumference, which is related to $\mathbf{u}_i$ by (Spadoni and Ruzzene, 2012)

$$\tilde{\mathbf{u}}_i = \mathbf{T}(\Theta_i)\mathbf{u}_i, \tag{26}$$

where the transformation matrix is

$$\mathbf{T}(\Theta_i) = \begin{pmatrix} 1 & 0 & -r\sin\Theta_i \\ 0 & 1 & \cos\Theta_i \\ 0 & 0 & 1 \end{pmatrix}, \tag{27}$$

with $\Theta_i$ being the azimuthal angle of the beam end on the circle (see Fig. 4). Giving the topological parameter $\beta$ and the azimuthal angle $\theta$ of the circle $i$ and $j$, the angle relations $\Theta_i = \pi/2 + \theta - \beta$ and $\Theta_j = 3\pi/2 + \theta - \beta$ are implied. DOFs of the beam ends $\tilde{\mathbf{u}}'$ in the local system ($\mathbf{e}_s - \mathbf{e}_n$ in Fig. 4) are then linked to the DOFs $\mathbf{u} = \{\mathbf{u}_i \;\; \mathbf{u}_j\}^T$ at the circle centers as

$$\tilde{\mathbf{u}}' = \mathbf{R}(\theta)\mathbf{T}(\Theta_1,\Theta_2)\mathbf{u}, \tag{28}$$

where

$$\mathbf{T}(\Theta_1,\Theta_2) = \begin{pmatrix} \mathbf{T}(\Theta_1) & 0 \\ 0 & \mathbf{T}(\Theta_2) \end{pmatrix}, \tag{29}$$

$$\mathbf{R}(\theta) = \begin{pmatrix} \mathbf{R}_{3\times3}(\theta) & 0 \\ 0 & \mathbf{R}_{3\times3}(\theta) \end{pmatrix}, \tag{30}$$

$$\mathbf{R}_{3\times3}(\theta) = \begin{pmatrix} \cos(\theta-\beta) & \sin(\theta-\beta) & 0 \\ -\sin(\theta-\beta) & \cos(\theta-\beta) & 0 \\ 0 & 0 & 1 \end{pmatrix}. \tag{31}$$

The stiffness of an Euler-Bernoulli beam in its local system is

$$\mathbf{K}' = \frac{E_s t}{6L} \begin{pmatrix} 1 & 0 & 0 & -1 & 0 & 0 \\ & 6t^2/L^2 & 3t^2/L^2 & 0 & -6t^2/L^2 & 3t^2/L^2 \\ & & 2t^2 & 0 & -3t^2/L^2 & t^2 \\ & & & 1 & 0 & 0 \\ & \text{sym} & & & 6t^2/L^2 & -3t^2/L^2 \\ & & & & & 2t^2 \end{pmatrix}. \tag{32}$$

where $E_s = E_b/(1-v_b^2)$ or $E_b$ under a plane strain or plane stress assumption, respectively,

with $E_b$ and $\nu_b$ being the Young's modulus and Poisson's ratio of the underlying ligament material. In a global system ($\mathbf{e}_x - \mathbf{e}_y$ in Fig.4), the stiffness matrix relating the general displacement and force vectors with respect to the center of the circles is obtained as

$$\mathbf{K} = \mathbf{T}^T(\Theta_i, \Theta_j)\mathbf{R}^T(\theta)\mathbf{K}'\mathbf{R}(\theta)\mathbf{T}(\Theta_i, \Theta_j). \tag{33}$$

Referring to Fig. 2, the strain energy of the three beams in a unit cell can be written as

$$w_1 = \frac{1}{2}\{\mathbf{u}_{p,q} \quad \mathbf{u}_{p+1,q}\}^T \mathbf{K}|_{\theta=0} \{\mathbf{u}_{p,q} \quad \mathbf{u}_{p+1,q}\} \tag{34a}$$

$$w_2 = \frac{1}{2}\{\mathbf{u}_{p,q} \quad \mathbf{u}_{p+1,q+1}\}^T \mathbf{K}|_{\theta=\pi/3} \{\mathbf{u}_{p,q} \quad \mathbf{u}_{p+1,q+1}\} \tag{34b}$$

$$w_3 = \frac{1}{2}\{\mathbf{u}_{p,q} \quad \mathbf{u}_{p,q+1}\}^T \mathbf{K}|_{\theta=2\pi/3} \{\mathbf{u}_{p,q} \quad \mathbf{u}_{p,q+1}\} \tag{34c}$$

respectively, and hence the strain energy of the unit cell $(p,q)$ is

$$w_{p,q}^{cell} = w_1 + w_2 + w_3. \tag{35}$$

We thus obtain the Hamiltonian of the whole system

$$H = \sum_{p,q}\left(w_{p,q}^{cell} + \frac{1}{2}m\dot{u}_{p,q}^2 + \frac{1}{2}m\dot{v}_{p,q}^2 + \frac{1}{2}J\dot{\phi}_{p,q}^2\right). \tag{36}$$

By using Hamilton's principle, the governing equations for the circle $\mathbf{u}_{p,q}$ are obtained as

$$\rho A_{cell} \frac{\partial^2 u_{p,q}}{\partial t^2} = -\frac{\partial H}{\partial u_{p,q}}, \tag{37a}$$

$$\rho A_{cell} \frac{\partial^2 v_{p,q}}{\partial t^2} = -\frac{\partial H}{\partial v_{p,q}}, \tag{37b}$$

$$j A_{cell} \frac{\partial^2 \phi_{p,q}}{\partial t^2} = -\frac{\partial H}{\partial \phi_{p,q}}, \tag{37c}$$

where the effective density and micro-inertia are defined as $\rho = m/A_{cell}$, $j = J/A_{cell}$, with $A_{cell} = \sqrt{3}a^2/2$ being the area of the unit cell. The right hand side of Eq.(37) contains DOFs $\mathbf{u}_{p,q}$ and those of its adjacent neighbors $\mathbf{u}_{p\pm1,q\pm1}$. The detailed expansion of Eq.(37) is straightforward but lengthy, and will not be presented here.

At long wave limit or the characteristic scale of the problem is much larger than the lattice constant $a$, the homogenization of the discrete Eq. (37) is possible by representing $\mathbf{u}_{p\pm1,q\pm1}$ as its Taylor series expansion at $\mathbf{u}_{p,q}$. Specifically, for the displacement $u$, we have

$$u_{p\pm1,q\pm1} = u + \mathbf{u}'^T d\mathbf{x}_{p\pm1,q\pm1} + \frac{1}{2} d\mathbf{x}^T_{p\pm1,q\pm1} \mathbf{u}'' d\mathbf{x}_{p\pm1,q\pm1} + O(|d\mathbf{x}_{p\pm1,q\pm1}|^2), \tag{38}$$

where

$$\mathbf{u}' = \{\partial u/\partial x \quad \partial u/\partial y\}^T, \tag{39a}$$

$$\mathbf{u}'' = \begin{pmatrix} \partial^2 u/\partial x^2 & \partial^2 u/\partial x \partial y \\ \partial^2 u/\partial x \partial y & \partial^2 u/\partial y^2 \end{pmatrix}, \tag{39b}$$

and from the relative sites of the circles (Fig. 2), the increments of the position vectors are given by

$$d\mathbf{x}_{p\pm1,q} = \{\pm a \quad 0\}^T, \tag{40a}$$

$$d\mathbf{x}_{p,q\pm1} = \{\mp a/2 \quad \pm\sqrt{3}a/2\}^T, \tag{40b}$$

$$d\mathbf{x}_{\pm(p+1,q+1)} = \{\pm a/2 \quad \pm\sqrt{3}a/2\}^T. \tag{40c}$$

Substituting Eq. (38) and the similar expansions of $v$ and $\phi$ into Eq.(37) and retaining the leading terms of the spatial differentials yield exactly the governing equations derived in the continuous description(Eq.17). This is an encouraging result which confirms the proposed chiral constitutive relation based on the continuum formulation. Then by equating the corresponding coefficients, the effective material constants of the equivalent chiral micropolar medium can be easily obtained,

$$\lambda = \frac{\sqrt{3}E_s}{4}\eta\left(\cos^2\beta - \eta^2\right)\sec^3\beta\cos2\beta, \tag{41a}$$

$$\mu = \frac{\sqrt{3}E_s}{4}\eta\left(\cos^2\beta + \eta^2\right)\sec^3\beta, \tag{41b}$$

$$\kappa = \frac{\sqrt{3}E_s}{2}\eta(\sin^2\beta + \eta^2)\sec\beta, \tag{41c}$$

$$\gamma = -\frac{E_s a^2}{4\sqrt{3}}\eta\left(3\sin^2\beta + 2\eta^2\right)\sec\beta, \tag{41d}$$

$$A = \frac{\sqrt{3}E_s}{2}\eta\left(\eta^2 - \cos^2\beta\right)\sec\beta\tan\beta, \tag{41e}$$

where all the constants are normalized as function of the ratio of the ligament wall thickness $(t \equiv t_b)$ to the lattice constant

$$\eta = t/a. \tag{42}$$

It can be verified that when the chiral lattice is flipped over (sign of $\beta$ is reversed), $A$ changes its sign, while all the other parameter remains unchanged, which is just the same as that predicted by Eq. (15) and (16). When $\beta = 0$ for a traditional triangular lattice, $A$ vanishes and the chiral effect disappears, the remaining effective material constants are reduced to

$$\lambda = \frac{\sqrt{3}E_s t}{4a^3}\left(a^2 - t^2\right), \tag{43a}$$

$$\mu = \frac{\sqrt{3}E_s t}{4a^3}\left(a^2 + t^2\right), \tag{43b}$$

$$\kappa = \frac{\sqrt{3}E_s t^3}{2a^3}, \tag{43c}$$

$$\gamma = -\frac{E_s t^3}{2\sqrt{3}a}, \tag{43d}$$

which are the same as those given by Kumar and McDonwell (2004) for a triangular lattice by considering up to the second order expansion of the microrotation DOF. With this homogenization scheme, the obtained higher order modulus $\gamma$ for the lattice structures is always negative, since the microration on the unit cell is approximated by a second order Taylor's expansion (Bazant and Christensen, 1972; Kumar and McDowell, 2004). Kumar and McDonwell (2004) also provided an insightful explanation on the negative sign of the higher order modulus $\gamma$. However, if the first order method (Chen et al, 1998; Kumar and McDowell, 2004; Spadoni and Ruzzene, 2012) is adopted, i.e. retaining the first order items of Eq. (38) and by using the equivalence of the unit cell energy density $w_{p,q}^{cell}/A_{cell}$ with the continuum

medium, Eq. (11), we can obtain a positive version of the higher order modulus as

$$\gamma = \frac{E_s a^2}{4\sqrt{3}} \eta \left(3\sin^2 \beta + 4\eta^2\right) \sec\beta. \tag{44}$$

while the other effective material constants in Eq. (41) remain unchanged. It also reduces to the results considering only the first order expansion when $\beta = 0$ (Chen et al, 1998; Kumar and McDowell, 2004).

## 4. Discussions and applications

In this section, we will discuss in detail the obtained effective material constants for the triangular chiral lattice. A static tension problem and a plane wave analysis will also be examined in order to illustrate the proposed theory.

*4.1 Properties of the effective material constants*

First, we will compare the derived effective material constants in Eq.(41) with those derived from a non-chiral micropolar theory (see Eq. (25-34) in Spadoni and Ruzzene (2012)) for the same chiral lattice. In the following, $\lambda_a, \mu_a, \kappa_a$ and $\gamma_a$ are denoted as the non-chiral version of the effective constants.

Comparing the results predicted by the chiral and non-chiral theories, it is found that the two theories give the same prediction on the modulus $\mu$ and $\kappa$ as well as the higher order modulus $\gamma$, but different on $\lambda$. Of course the parameter $A$ is only introduced in the current chiral theory. In Fig. 5, $A$, $\lambda$ and $\lambda_a$ normalized by $E_s$ are plotted as function of $\beta$, where $\eta = 1/20$ is used in the calculation, it is shown that the chiral constant is an odd function of $\beta$ and $|A|$ increases with the increase of $|\beta|$ in the region of roughly $|\beta| < 75°$, this indicates that the bulk-rotation coupling behavior is stronger with the increasing circle size. It should also be noted here that since the beam theory is used the results may not be accurate or even meaningless for $\beta \to 90°$, where the ligament is vanishingly short. An additional curve is also included in the figure for the parameter $A$ with a slenderness $\eta = 1/50$, the result shows that the chiral constant also increases for greater ligament width. Physically positive $A$ means, in the constitutive equation Eq.(15), an anticlockwise circle rotation, which will induce a hydrostatic compression, while all the other strain components are hold to be zero. This is just the case for the lattice shown in Fig. 3a. The scenario should be reversed for another handedness, Fig.3b. Finally, from Eq. (41), we get

$$(\lambda + \mu)\kappa - A^2 = \frac{3}{4}(\eta \sec \beta)^4 \geq 0, \tag{45}$$

which confirms that the thermodynamic stability requirement on the chiral constant, Eq. (22), is always fulfilled.

An unusual characteristic of the non-chiral micropolar (and also the classical elasticity since they are the same in bulk behavior) homogenization of the chiral lattice is the negativity of $\lambda$. This is obvious since the chiral lattice produces negative Poisson's ratio $\nu = \lambda/(\lambda + 2\mu)$. However, Fig. 5 shows that the current chiral homogenization gives significant variation of $\lambda$, thus it seems that the effective Poisson's ratio would be no longer negative in some ranges of the parameter $\beta$. Based on the constitutive equations Eqs. (15) and (16), the Poisson's ratio of a chiral micropolar medium must be redefined. By definition, $\nu$ is defined as $-\varepsilon_{22}/\varepsilon_{11}$, assuming all the stress components except for $\sigma_{11}$ are zero, we have

$$\lambda \varepsilon_{11} + (\lambda + 2\mu)\varepsilon_{22} - A(\varepsilon_{12} - \varepsilon_{21}) = 0 \tag{46a}$$

$$-A(\varepsilon_{11} + \varepsilon_{22}) + (\mu + \kappa)\varepsilon_{12} + (\mu - \kappa)\varepsilon_{21} = 0 \tag{46b}$$

$$A(\varepsilon_{11} + \varepsilon_{22}) + (\mu - \kappa)\varepsilon_{12} + (\mu + \kappa)\varepsilon_{21} = 0 \tag{46c}$$

Solving the above equations gives the Poisson's ratio for the chiral microplar medium

$$\nu = \frac{\lambda - A^2/\kappa}{\lambda + 2\mu - A^2/\kappa}. \tag{47}$$

The sign of the chiral constant does not affect the Poisson's ratio. Physically this means the auxetic behavior is independent of the direction of internal rotation of the circles, as expected.

The Young's modulus now becomes

$$E = \frac{(\lambda + 2\mu - A^2/\kappa)^2 - (\lambda - A^2/\kappa)^2}{\lambda + 2\mu - A^2/\kappa}. \tag{48}$$

By using Eq.(41), the overall Young's modulus and Poisson's ratio of the chiral lattice are explicitly expressed as

$$v = \frac{1 - \eta^2 - (\cos\beta\sin\beta/\eta)^2}{3 + \eta^2 + (\cos\beta\sin\beta/\eta)^2}. \tag{49}$$

$$E = 2\sqrt{3}E_s\eta^3 \frac{\sec^3\beta\left[\eta^2 + \cos^2\beta\right]}{3\eta^2 + \eta^4/8 + \cos^2\beta\sin^2\beta}. \tag{50}$$

From Eq.(49), $v \approx 1/3$ (slightly affected by the ligament slenderness) for the two extreme values of $\beta$, while for the other geometry $v \approx -1$ since the $1/\eta$ term dominates. In fact, it is interesting to verify that the above expressions are exactly the same as the non-chiral prediction by Spadoni and Ruzzene (2012). The other commonly used parameters such as the characteristic length $l^2 = \gamma/\mu$ and the coupling number $N^2 = \kappa/(\kappa + \mu)$ (Cowin, 1970) are also the same as those derived from the non-chiral theory. In summary, the proposed chiral micropolar theory introduces the chiral parameter $A$ to capture the coupling between the bulk deformation and the internal rotation, and is only different in the prediction on Lame's constant $\lambda$ compared with the non-chiral micropolar theory

*4.2 One dimensional static tension*

We consider in the following a one dimensional tension problem of the chiral micropolar medium, as shown in Fig. 6a. A sourceless domain is $l$ in size in $x$ direction, and infinite in $y$ direction. A constant displacement $u^o$ is prescribed on the right side while all the other DOFs are fixed on the boundary. Since the problem is infinite in $y$ direction, the field quantities are only the functions of $x$, the strains are then $\varepsilon_{xx} = u'$, $\varepsilon_{yy} = 0$, $\varepsilon_{xy} = v' - \phi$ and $\varepsilon_{yx} = \phi$, where a prime means the ordinary differential with respect to $x$. With the help of the constitutive and equilibrium equations, we obtain

$$(\lambda+2\mu)u''-Av''+2A\phi'=0, \tag{51a}$$

$$-Au''+(\mu+\kappa)v''-2\kappa\phi'=0, \tag{51b}$$

$$\gamma\phi''-4\kappa\phi+2\kappa v'-2Au'=0, \tag{51c}$$

with the boundary conditions shown in Fig.6a.

For the classical micropolar case ($A=0$), the solution is identical with that by the Cauchy elasticity, since Eq. (51a) is decoupled from the other two and forms itself a complete boundary problem, while $v$ and $\phi$ are zero conforming the homogeneous boundary conditions. When the chirality appears, the variables in Eq. (51) are coupled together, this implies that even for a simple tension problem in $x$ direction will produce the displacement in $y$ direction, as well as the rotational field. This basic behavior cannot be predicted with the existing 2D version of the non-centrosymmetric micropolar theory.

As an example, Eq. (51) is solved with the parameters $l=20a$, $\eta=1/20$, and $u^o=a/10$. A discrete model of the finite element method (FEM) is constructed with the same parameters as sketched in Fig.6b. The rigid circle is modeled by the constraint equations which connect the six nodes on the circle to the nodes at its center. The infinite size in $y$ direction is mimicked by periodic boundary conditions with the matched nodes on each side, and the same boundary conditions as Fig. 6a are point-wisely enforced. The displacement $v$ and the microrotation $\phi$ predicted by the analytical and FEM solutions are displayed in Fig. 6c, a fairly well agreement is found between the two methods. Due to the chirality of the lattice structure, the displacement in $y$ increases in an asymmetrical manner in the vicinity of the boundary, and soon becomes affine across the $x$ length, which creates an inclined deformed sample. It is also found that the significant variation portion of the fields near the boundary is nearly independent of the length $l$, this can be attributed to the boundary layer effect (Eringen, 1999) and agrees with the traditional micropolar theory. If the handedness of the microstructure of the FEM model is changed, i.e., the model is mirror reflected with respect to $x$ axis, the field distribution should be also reflected. For the analytical model, this is a natural result by inversing the sign of the chiral constant ($-A$). The non-chiral micropolar theory always gives trivial solutions for $v$ and $\phi$. Though the displacement in $y$ caused by the tension in $x$ is relatively small for the static case, such effect would become more pronounced for wave propagation problems demonstrated as follows.

*4.3 Plane wave propagation*

Considering in an infinite planar chiral micropolar medium under a plane wave in $+x$ direction, the displacement and the microrotation are assumed to be the following from

$$(u, v, \phi) = (\hat{u}, \hat{v}, \hat{\phi}) \exp(ikx - i\omega t), \tag{52}$$

where $k$ and $\omega$ denote the wave number and circular frequency, $(\hat{u}, \hat{v}, \hat{\phi})$ are the (complex) amplitudes, and $i = \sqrt{-1}$. Substituting Eq. (52) into (17) yields the following secular equation system for the chiral micropolar medium.

$$\begin{pmatrix} k^2(\lambda + 2\mu) - \omega^2 \rho & -Ak^2 & -2iAk \\ -Ak^2 & k^2(\kappa + \mu) - \omega^2 \rho & 2i\kappa k \\ 2iAk & -2ik\kappa & (k^2\gamma + 4\kappa) - \omega^2 j \end{pmatrix} \begin{pmatrix} \hat{u} \\ \hat{v} \\ \hat{\phi} \end{pmatrix} = 0. \tag{53}$$

Let $A = 0$, we obtain the secular equation for a classical micropolar medium without chirality

$$\begin{pmatrix} k^2(\lambda + 2\mu) - \omega^2 \rho & 0 & 0 \\ 0 & k^2(\kappa + \mu) - \omega^2 \rho & 2i\kappa k \\ 0 & -2ik\kappa & (k^2\gamma + 4\kappa) - \omega^2 j \end{pmatrix} \begin{pmatrix} \hat{u} \\ \hat{v} \\ \hat{\phi} \end{pmatrix} = 0. \tag{54}$$

The most pronounced difference between the chiral and non-chiral micropolar media is that for the later one a non-dispersive longitudinal wave with velocity $c_p = \sqrt{(\lambda + 2\mu)/\rho}$ can always be decoupled from the other two shear-rotation coupled waves (Eringen, 1999). This is the characteristic of the non-chiral micropolar media, i.e. the microrotation is only coupled with shear but not with dilatation. In the current chiral micropolar theory, as implied by Eqs.(11-13), the rotation is coupled with the dilation deformation due to the non-zero chiral constant $A$. Hence there would be no longer pure P or pure S wave in such media. The three wave modes are all mixed and dispersive, thus we call P, S or R (rotation) dominated waves, respectively. Since the full matrix in Eq.(53) will result in a full third order equation, the analytical eigen-solution is tedious, we will instead numerically evaluate it through examples.

The wave behavior in an isotropic 3D chiral micropolar medium has been addressed in the literature (Lakhtakia et al, 1988; Ro, 1999; Khurana and Tomar, 2009). A remarkable feature is that the transverse (shear and rotation coupled) waves can be distinguished as left circular polarized (LCP) and right circular polarized (RCP) waves, which means that the ratio of the two displacement (also the rotation) components equals to $\pm i$ and not in the same phase. The LCP and RCP waves propagate at different phase speeds, this reveals the lack of mirror symmetry of the underlying chiral microstructure inherent in the material. From Eq.(53), it is easy to obtain the ratio of the two displacement amplitudes as

$$\frac{\hat{u}}{\hat{v}} = \frac{k^2[A^2 - \kappa(\lambda + 2\mu)] + \kappa\omega^2\rho}{A(k^2\mu - \omega^2\rho)}. \tag{55}$$

Obviously this ratio cannot be an imaginary number, therefore the common feature (circular polarization) for 3D chiral micropolar media is not presented in the 2D version, and the particle motion under the wave is always linearly polarized. However, the loss of mirror symmetry is reflected in another way for the 2D case. Because the medium is in-plane isotropic, the frequency dispersion must be isotropic and independent of the propagating direction, i.e. the iso-frequency contour of this medium should be concentric circles. On the other hand, the displacement polarization will remain the same angle with respect to the wave vector. This result is schematically shown in Fig. 7a. The polarization according to the wave vector **k** forms a pattern with a rotational symmetry, but without mirror reflective symmetry. It is interesting to note that the P- and S- mixed wave modes can be traditionally observed in the anisotropic medium, but we can find the mixed wave modes in a 2D chiral micropolar medium while the dispersion is isotropic. This wave phenomenon is not reported before.

In the following, numerical solutions based on the chiral and non-chiral micropolar theories for the same chiral lattice will be conducted. For comparison, the exact dispersive solution of the chiral lattice is also calculated from the Bloch wave (Brillouin, 1953) solution of the discrete kinetic equation (37), by adopting the following lattice wave function

$$(u,v,\phi)_{p,q} = (\hat{u},\hat{v},\hat{\phi})_{p,q} \exp(ikx_{p,q} - i\omega t) \tag{55}$$

where the location of certain rigid circle $x_{p,q}$ takes the similar value as Eq. (40). The wave frequency $\omega$ is normalized by

$$\Omega = \sqrt{\frac{4E_s t^3}{ma^3}} \tag{56}$$

which represents the natural frequency of a simple supported beam with a lumped mass at its middle. Parameters $\beta = 0.9$, $\eta = 1/20$, $m = J = 1$ are used in the numerical example.

Fig. 7b shows the three branches (in black, red and blue, respectively) of the dispersion curves, predicted by the chiral micropolar theory (solid line), the non-chiral micropolar theory (dashed line) and the discrete model (circles), respectively. The first and second branches correspond to the displacement dominated modes and the third one is the rotational dominated wave. The second branch is almost non-dispersive. For the non-chiral theory this is just the uncoupled non-dispersive P-wave, while they are slightly dispersive predicted by both the chiral theory and discrete model. All three models agree well for this branch. However, for the first and third branches, the chiral theory agrees well with those given by the discrete model and a large discrepancy between the chiral and the non-chiral micropolar theory is found.

To further examine the wave modes, here we focus on the displacement dominated one, i.e., the first and second branches. Since the particle is linearly polarized, it is convenient to define a polarization angle $\Lambda$ with respect to the wave propagating direction, i.e. in $x$ direction

$$\tan \Lambda = |\hat{v}/\hat{u}|. \tag{57}$$

The polarization angle of the two waves predicted by the chiral and non-chiral theories are plotted in Fig.7c as function of the wave number, where the first and second branches are marked in black and red, respectively,. The non-chiral theory always predicts pure P and S-waves, thus the polarization angles remain to be 0 and 90 degrees, as expected. For the chiral theory, the S and P dominated wave (for example $\Lambda > 60°$ and $\Lambda < 30°$) can be observed when the wave number is small ($ka < 0.05$) for the first and second branches, respectively. However, for the intermediate wave number ($ka > 0.05$), the first two branches become indistinguishable and can even interchange, i.e. the first branch become P dominated and the second branch become S dominated. Again, the predictions by the proposed chiral micropolar model always agree well with those by the discrete model (shown by circle and square dots). Polarizations corresponding to different topology parameters are also shown in Fig. 7d. It is found that the P- and S- wave modes will be more mixed for larger $\cos\beta$, i.e., smaller circles for the same lattice constant.

## 5. Conclusions

The existing micropolar theory is not able to characterize the chiral effect inherent in plane isotropic chiral solids, e.g. triangle chiral lattices, to this end, we propose in this work a continuum theory to capture the chiral effect in such materials. The proposed method is based on the micropolar theory and reinterpretation of isotropic tensors with in-plane isotropy. The constitutive equation and the governing equation are analytically derived. Different from the existing 2D isotropic micropolar theory, a new material constant is introduced to characterize the chiral effect, it can be either positive or negative depending on the handedness of the material while its absolute value is bounded. Physically, the proposed chiral micropolar theory can describe a dilatation-rotation coupling, in addition to the shear-rotation coupling by the traditional micropolar elasticity.

We also propose a homogenization scheme for a triangle chiral lattice, from which the effective material parameters of the chiral lattice are derived analytically. By introducing the new material parameter, the proposed homogenization method gives different prediction on the Lames constant $\lambda$, but the other effective material constants remain the same as those predicted by the non-chiral version. With the proposed theory, we also examine a static tension and a plane wave loading cases for a triangle chiral lattice. In the static tension loading, it is found that even a simple tension can induce rotation and lateral displacement fields inside of the material, as expected. For the plane wave loading, there is no longer pure longitudinal and transverse waves, and all the wave modes are of mixed types and dispersive. Mixed wave modes with isotropic frequency dispersion are also observed, which is a fundamental property of a planar isotropic chiral solid. This special wave phenomenon is not reported before. These results are also confirmed by the exact solution of the corresponding discrete model. The proposed method provides a useful tool to investigate the chiral effect on mechanical behavior for plane isotropic chiral solids.

**Acknowledgment**

This work was supported in part by Air Force Office of Scientific Research under Grant No. AF 9550-10-0061 with Program Manager Dr. Byung-Lip (Les) Lee and NSF 1037569, and in part by National Natural Science Foundation of China under Grants No. 10972036,10832002 and 11072031.

# References

Alderson, A., Alderson, K.L., Attard, D., Evans, K.E., Gatt, R., Grima, J.N., Miller, W., Ravirala, N., Smith, C.W., Zied, K., 2010. Elastic constants of 3-, 4- and 6-connected chiral and anti-chiral honeycombs subject to uniaxial in-plane loading. Compos. Sci. Technol. 70, 1042-1048.

Bazant, Z.P., Christensen, M., 1972. Analogy between micropolar continuum and grid frameworks under initial stress. Int. J. Solids Struct. 8, 327-346.

Borisenko, A.I., Tarapov, I.E., 1979. Vector and Tensor Analysis with Applications. Dover Publications, New York.

Brillouin, L., 1953. Wave Propagation in Periodic Structures. Dover, New York.

Chandraseker, K., Mukherjee, S., 2006. Coupling of extension and twist in single-walled carbon nanotubes. J. Appl. Mech. 73, 315-326.

Chandraseker, K., Mukherjee, S., Paci, J.T., Schatz, G.C., 2009. An atomistic-continuum Cosserat rod model of carbon nanotubes. J. Mech. Phys. Solids 57, 932-958.

Chen, J.Y., Huang, Y., Ortiz, M., 1998. Fracture analysis of cellular materials: a strain gradient model. J. Mech. Phys. Solids 46, 789-828.

Cosserat, E., Cosserat, F., 1909. Théorie des Corps Déformables. A. Hermann, Paris.

Cowin, S.C., 1970. An incorrect inequality in micropolar elasticity theory. Z. Angew. Math. Phys. 21, 494-497.

Dirrenberger, J., Forest, S., Jeulin, D., Colin, C., 2011. Homogenization of periodic auxetic materials. Procedia Engineering 10, 1847-1852.

Eringen, A.C., 1966. Linear theory of micropolar elasticity. J. Math. Mech. 15, 909-923.

Eringen, A.C., 1999. Microcontinuum Field Theories I: Foundations and Solids. Springer, New York.

Ieşan, D., 2010. Chiral effects in uniformly loaded rods. J. Mech. Phys. Solids 58, 1272-1285.

Joumaa, H., Ostoja-Starzewski, M., 2011. Stress and couple-stress invariance in non-centrosymmetric micropolar planar elasticity. Proc. R. Soc. A doi:10.1098/rspa.2010.0660 Published online.

Khurana, A., Tomar, S.K., 2009. Longitudinal wave response of a chiral slab interposed between micropolar solid half-spaces. Int. J. Solids Struct. 46, 135-150.

Kelvin, Lord, 1904. Baltimore Lectures on Molecular Dynamics and the Wave Theory of Light. C. J. Clay and Sons, London.

Kumar, R.S., McDowell, D.L., 2004. Generalized continuum modeling of 2-D periodic cellular solids. Int. J. Solids Struct. 41, 7399-7422.

Lakes, R.S., Benedict, R.L., 1982. Noncentrosymmetry in micropolar elasticity. Int. J. Eng. Sci. 20, 1161-1167.

Lakes, R.., Yoon, H.S., Katz, J.L., 1983. Slow compressional wave propagation in wet human and bovine cortical bone. Science 220, 513-515.

Lakes, R.S., 1987. Foam structures with a negative Poisson's ratio. Science 235, 1038-1040.

Lakes, R., 2001. Elastic and viscoelastic behavior of chiral materials. Int. J. Mech. Sci. 43, 1579-1589.

Lakhtakia, A., Varadan, V.V., Varadan, V.K., 1988. Elastic wave propagation in noncentrosymmetric, isotropic media: Dispersion and field equations. J. Appl. Phys. 63, 5246-5250 .

Liu, X.N., Hu, G.K., 2005. A continuum micromechanical theory of overall plasticity for particulate composites including particle size effect, Int. Journal of Plasticity. 21, 777-799.

Liu, X.N., Hu, G.K., Sun, C.T., Huang, G.L., 2011a. Wave propagation characterization and design of two-dimensional elastic chiral metacomposite. J. Sound Vib. 330, 2536-2553.

Liu, X.N., Hu, G.K., Huang, G.L., Sun, C.T., 2011b. An elastic metamaterial with simultaneously negative mass density and bulk modulus. Appl. Phys. Lett. 98, 251907.

Natroshvili, D., Stratis, I.G., 2006a. Mathematical problems of the theory of elasticity of chiral materials for Lipschitz domains. Math. Methods Appl. Sci. 29, 445-478.

Natroshvili, D., Giorgashvili, L., Stratis, I.G., 2006b. Representation formulae of general solutions in the theory of hemitropic elasticity. Q. J. Mech. Appl. Math. 59, 451-474.

Nowacki, W., 1986. Theory of Asymmetric Elasticity. Pergamon Press, New York.

Prall, D., Lakes, R.S., 1996. Properties of a chiral honeycomb with a Poisson's ratio $\approx$ -1. Int. J. Mech. Sci. 39, 305-314.

Ro, R., 1999. Elastic activity of the chiral medium. J. Appl. Phys. 85, 2508–2513.

Spadoni, A., Ruzzene, M., Gonella, S., Scarpa, F., 2009. Phononic properties of hexagonal chiral lattices. Wave Motion 46, 435-450.

Spadoni, A., Ruzzene, M., 2012. Elasto-static micropolar behavior of a chiral auxetic lattice. J. Mech. Phys. Solids 60, 156-171.

Wang, J.S., Feng X. Q., ,X. Jun, Qin Q.H.,Yu, S.W., 2011. Chirality transfer from molecular to morphological scales in quasi-one-dimensional nanomaterials: a continuum model, J. of Computational and Theoretical Nanoscience 8, 1-10.

Yang, W., Li, Z.M., Shi, W., Xie, B.H., Yang, M.B., 2004. Review on auxetic materials. J. Mater. Sci. 39, 3269-3279.

**Figure captions**

Fig. 1 Visualization of the physical meaning of the 2D chiral micropolar constitutive equation.

Fig. 2 Geometry of auxectic triangular chiral lattice.

Fig. 3 Chiral lattice with (a) $\beta > 0$ and its handedness reversed pattern with (b) $\beta < 0$.

Fig. 4 Geometry and symbol definitions for the analysis of a single ligament.

Fig. 5 Effective Lame constant $\lambda$ predicted by chiral and non-chiral theory, and the effective chiral constant with different microstructure paramerters.

Fig. 6 (a) Problem definition and (b) sketch of corresponding discrete FEM model for the one-dimensional tension; (c) lateral displacement and microrotaiton fields.

Fig. 7 (a) Schematic wave behavior of the 2D chiral media; Comparison of (b) dispersion relation and (c) polarization angle for the chiral, non-chiral homogenization and the discrete model; (d) Polarization variation for the first two branches with different topology parameters.

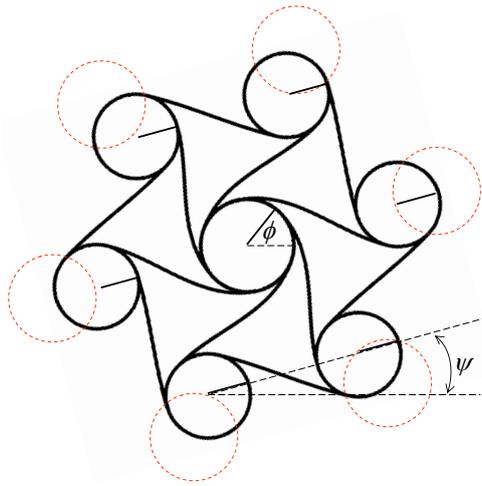

Fig. 1

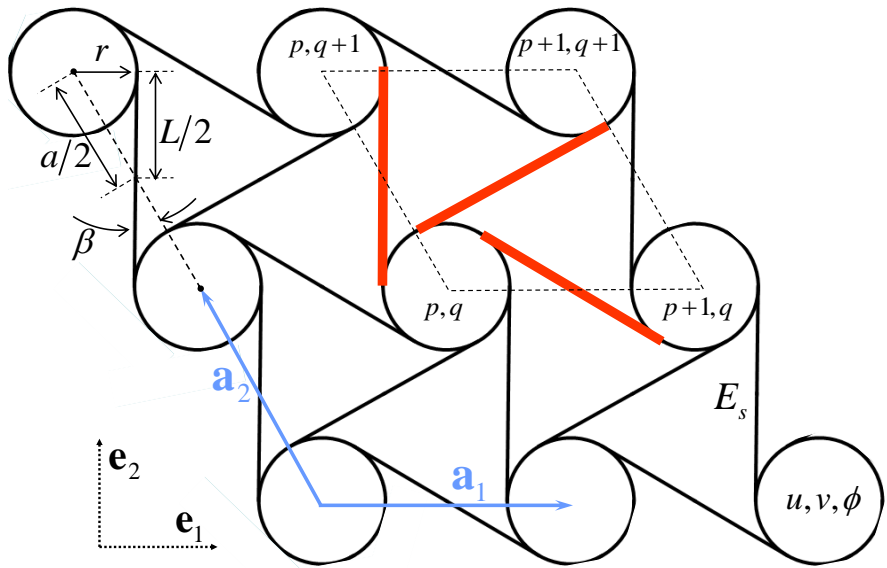

Fig. 2.

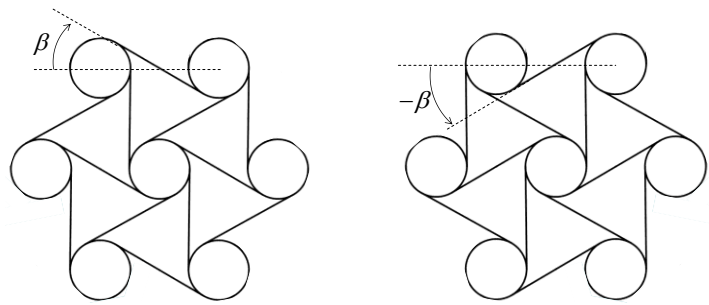

Fig. 3.

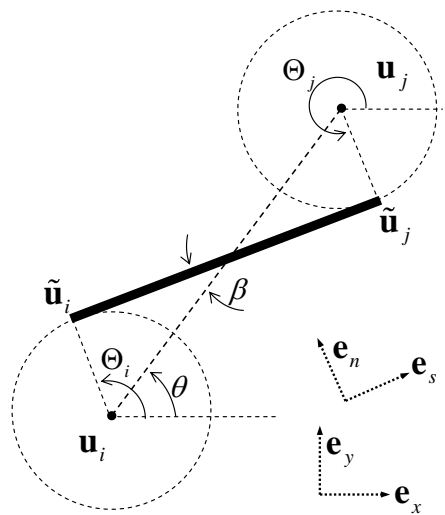

Fig. 4

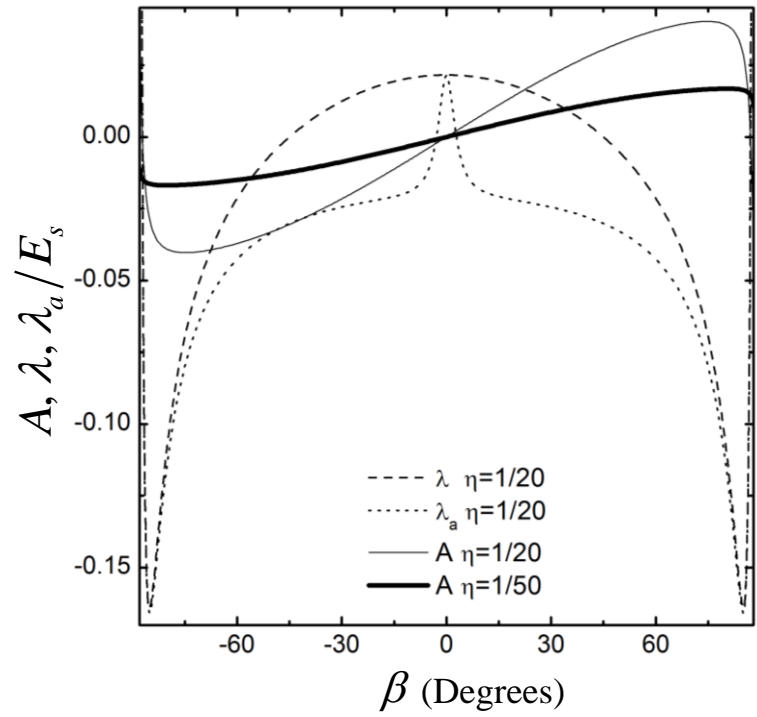

Fig. 5

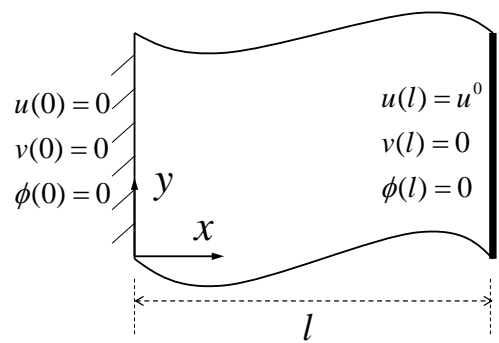

Fig. 6a

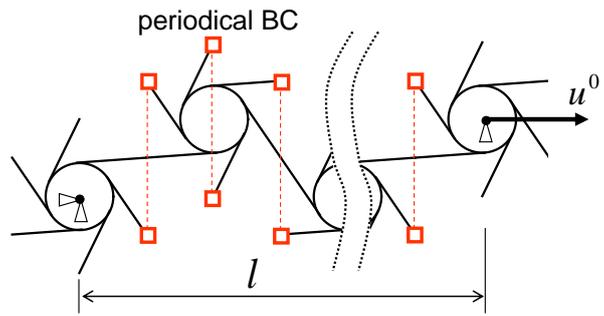

Fig. 6b

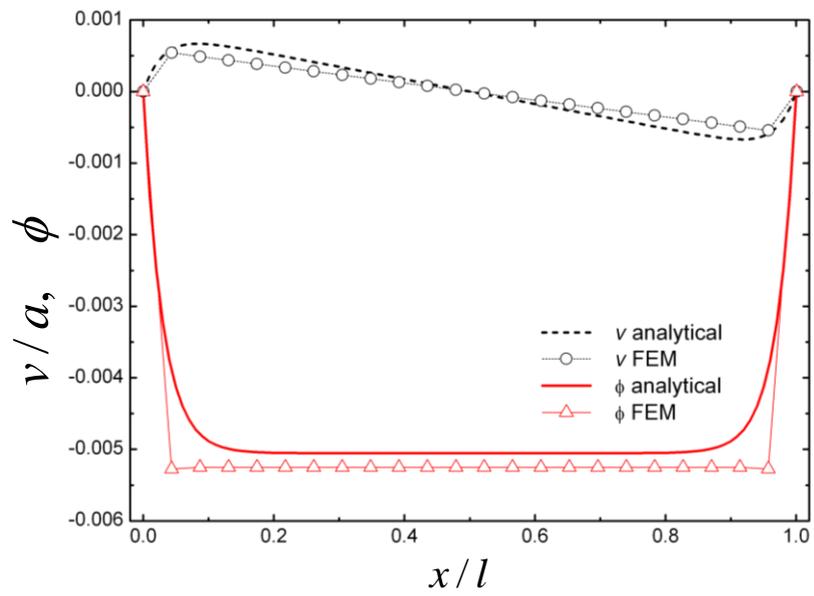

Fig. 6c

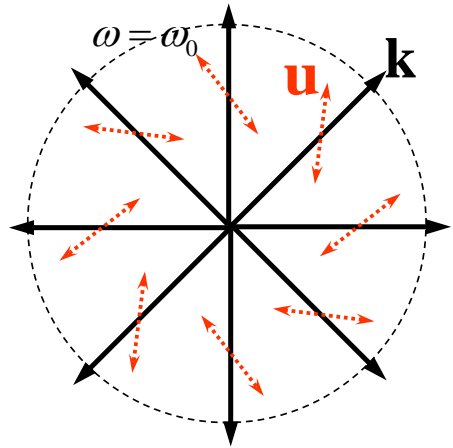

Fig. 7a

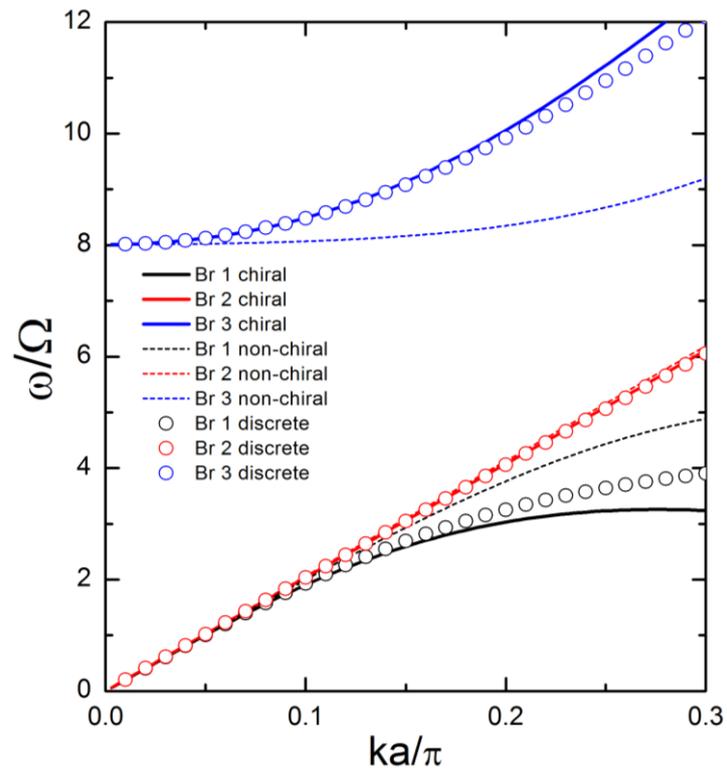

Fig. 7b

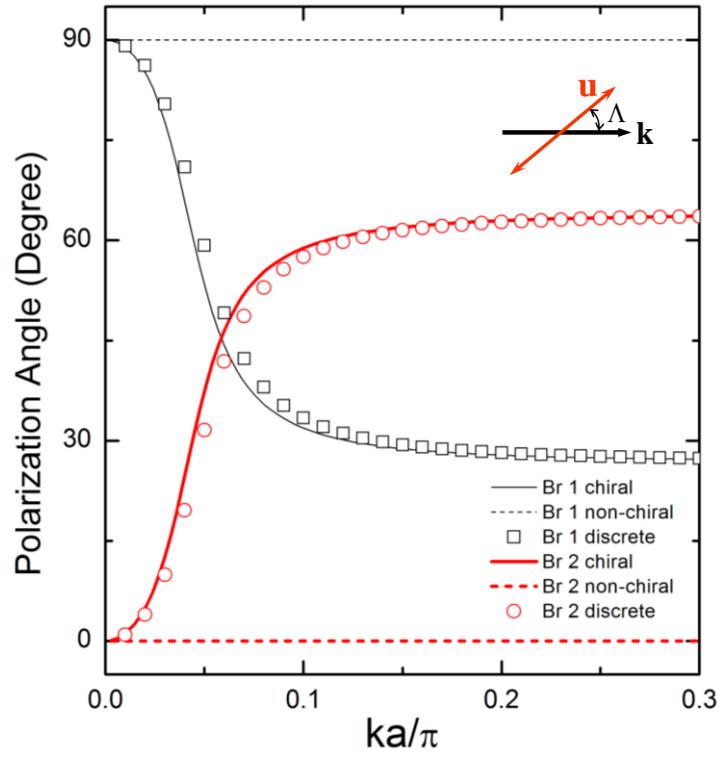

Fig. 7c

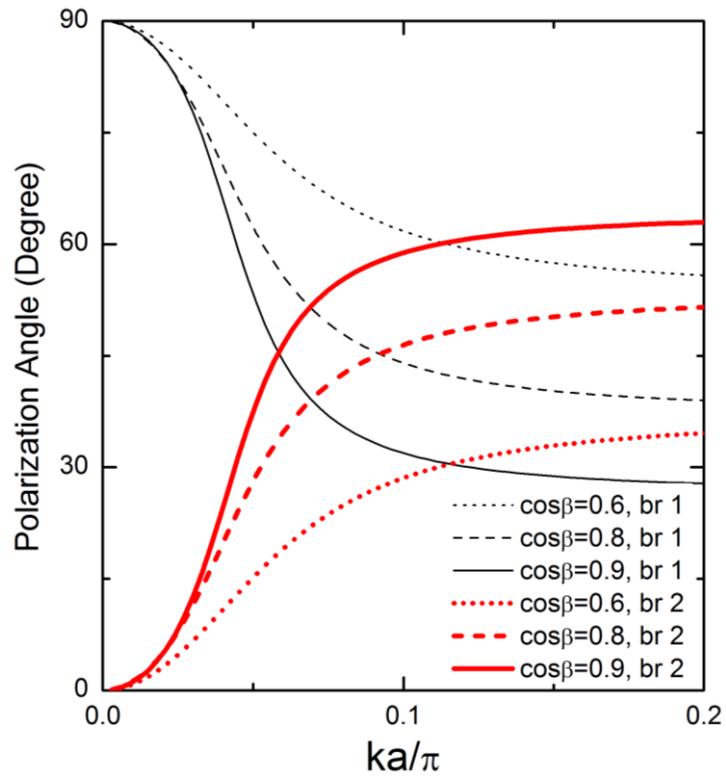

Fig. 7d